\documentclass[aps,prl,twocolumn,groupedaddress,showpacs]{revtex4}

\usepackage{amssymb,amsmath,graphicx}

\begin{document}

\title{Magnetically Stimulated Diffusion of Rydberg Gases}

\author{Yurii V. Dumin}
\email[]{dumin@yahoo.com}
\affiliation{
Moscow State University,
GAISh, Universitetski pr.\ 13, 119992, Moscow, Russia}
\affiliation{
Space Research Institute (IKI) of Russian Academy of Sciences,
Profsoyuznaya str.\ 84/32, 117997, Moscow, Russia}
\altaffiliation{On leave from:
Theoretical Department, IZMIRAN, Russian Academy of Sciences,
Troitsk, Moscow reg., 142190 Russia
}

\date{November 26, 2012}

\begin{abstract}
The specific kind of diffusion stimulated (rather than suppressed)
by the external magnetic field, which was predicted for the first
time by Schmelcher and Cederbaum in 1992, is considered here for
the case of high-angular-momentum (\textit{i.e.}, approximately
``circular'') Rydberg atoms. The coefficient of such diffusion was
calculated by a purely analytical approach and was found to be
well relevant to the experiments on antihydrogen formation.
\end{abstract}

%

\maketitle

A well-known feature of the diffusion processes in various
gaseous systems (ranging from the laboratory devices for
plasma confinement to the large-scale astrophysical objects)
is that it is strongly suppressed by imposition of the external
magnetic fields.
The aim of the present article is to discuss the case when Rydberg
atoms, along with their other extraordinary properties (\textit{e.g.},
review~\cite{gal94}), may exhibit also a quite unexpected kind of
diffusion, which is stimulated rather than suppressed by
the magnetic field.

In fact, the possibility of such phenomenon was mentioned
for the first time by Schmelcher and Cederbaum as early as
1992~\cite{sch92a,sch92b}. These authors, using a classical
approximation for the description of hydrogen-like atom in
a magnetic field, studied dynamics of a ``cyclic'' canonical coordinate
(\textit{i.e.}, the one not entering explicitly into the Hamiltonian),
such as the center of mass of the atom. As a result, they revealed
a possibility of chaos development in this coordinate and found that,
in some circumstances, such chaotic motion follows the classical
diffusion law.

Later, a profound mathematical treatment of the same problem
was given in paper~\cite{bab06}. Its authors used the method of
separation between the ``slow'' and ``fast'' degrees of freedom
of a dynamical system. Then, the fast chaotic motions were
treated as a stochastic force, resulting in the diffusion-like
behavior of the slow degrees of freedom.

Experimental observation of such diffusion requires either
an extremely strong magnetic field or very weak interatomic
electric field (\textit{i.e.}, a low binding energy).
Therefore, it was suggested to observe this phenomenon either
in compact astrophysical objects with ultrastrong magnetic
fields (\textit{e.g.}, pulsars) or in the experiments with
highly-excited (Rydberg) atoms. Unfortunately, none of these
opportunities was realized by now because, on the one hand,
the diagnostic possibilities in astrophysical studies are
too limited and, on the other hand, the standard atomic-beam
experiments with Rydberg atoms are not well suited for
studying the diffusion processes.

Fortunately, the situation changed in the recent decade:
magneto-optical traps (MOT) became a new source of
the Rydberg atoms, which are formed due to recombination
of laser-produced ultracold plasmas
(\textit{e.g.}, reviews~\cite{gou01,ber03});
and such devices are much better suited for tracing
the long-term diffusional effects.
Besides, quite similar apparatus is used now to create and
study the antihydrogen by recombining antiprotons and
positrons~\cite{rei10,and10}. Such antihydrogen atoms are also
born in the highly-excited states; and strong magnetic fields
are applied to prevent antiparticles from the escape and
annihilation.

However, it should be kept in mind that the above-cited
theoretical studies were concentrated mostly on the case of
atoms with zero or very low angular momenta.
This was quite natural for the old atomic-beam experiments with
Rydberg atoms produced by laser irradiation: such atoms can have
a huge principal quantum number, but their orbital number is always
limited to a few units, because any absorbed photon can change
it only by unity.
On the other hand, the Rydberg atoms formed by the recombination
in MOT experiments, in general, should have large angular momenta
(so that the respective orbital quantum numbers are on the order of
the principal number). It is not clear in advance if such atoms will
behave similarly in the external magnetic fields. So, a special
treatment of the high-orbital-momentum case should be performed.
Besides, to estimate importance of the magnetically-stimulated
diffusion in various experimental setups, it would be desirable
to have an analytical expression for the diffusion coefficient,
with explicit dependences on all the relevant parameters.
It is just the aim of the present work to provide such formula
\footnote{
Let us mention also one more subtle point of the earlier
theoretical studies: it is not clear if the classical equations of
motion have a real physical sense for the Rydberg atoms with very
low angular momenta because, generally speaking, \textit{all}
quantum numbers should be large for the classical mechanics
to work well. Fortunately, the case of high-angular-momentum
Rydberg atoms considered in our paper satisfies this requirement.}.

\begin{figure}[t]
\includegraphics[width=4.8cm]{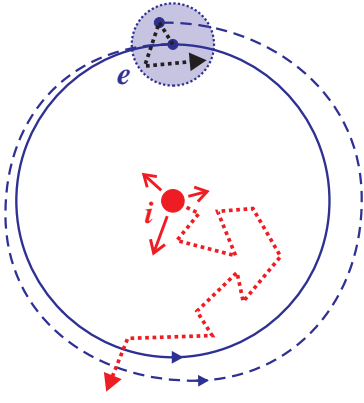}
\caption{\label{fig:Orbits}
Sketch of development of the diffusion-like behavior of
ion (\textit{i}) under the action of revolutions of
the electron (\textit{e}) perturbed by the external magnetic
field.}
\end{figure}

Before proceeding to the detailed quantitative treatment,
let us try to explain qualitatively why the magnetic field can
stimulate rather than suppress a diffusion. In the case of
high-angular-momentum (\textit{i.e.}, approximately ``circular'')
Rydberg atoms, this can be presented pictorially in
Fig.~\ref{fig:Orbits}.
Indeed, if the magnetic field is absent, the electron orbit will
be exactly closed, and average force experienced by the central
ion after each revolution of the electron will equal exactly zero.
On the other hand, when an external magnetic field is imposed
on the system, then the curvature radius of the electron orbit will
either increase or decrease under the action of the additional
Lorentz force. As a result, the orbit will no longer be closed;
and a position of the electron after each revolution will be
slightly shifted with respect to the initial point, as illustrated
by the shaded circle in Fig.~\ref{fig:Orbits}. Therefore,
the average Coulomb force acting on the ion after such revolution
becomes nonzero; and this ion will experience a kick in
a quasi-random direction. A series of such kicks is equivalent
to a random force, leading to the diffusion-like Brownian
motion of the ion and, consequently, of the entire atom.

Quantitative description of the two-body system, coupled by
Coulomb forces and imbedded in the external magnetic field,
can be evidently given by the set of equations:
\begin{equation}
\left\{ \!
\begin{array}{l}
m_e \frac{\displaystyle d^2}{\displaystyle d t^2} \, {\rm \bf r}_e =
  - e^2 \frac{\displaystyle {\rm \bf r}_e \! - {\rm \bf r}_i }%
  {\displaystyle | \, {\rm \bf r}_e \! - {\rm \bf r}_i |^3 } \, - \,
  \frac{\displaystyle \vphantom{l_x^x} e}{\displaystyle \vphantom{l_x^x} c} \,
  \frac{\displaystyle d {\rm \bf r}_e}{\displaystyle d t}
  \! \times \! {\rm \bf B} \, ,
\\[3ex]
m_i \, \frac{\displaystyle d^2}{\displaystyle d t^2} \, {\rm \bf r}_i \, =
  \:\:\: e^2 \frac{\displaystyle {\rm \bf r}_e \! - {\rm \bf r}_i }%
  {\displaystyle | \, {\rm \bf r}_e \! - {\rm \bf r}_i |^3 } \: + \:
  \frac{\displaystyle \vphantom{l_x^x} e}{\displaystyle \vphantom{l_x^x} c} \,
  \frac{\displaystyle d {\rm \bf r}_i}{\displaystyle d t}
  \! \times \! {\rm \bf B} \, ,
\end{array}
\right.
\label{eq:Initial_set}
\end{equation}
where $ m_e $ and $ m_i $ are the electron and ion masses,
$ {\rm \bf r}_e $ and $ {\rm \bf r}_i $ are their radius vectors,
$ e $~is the absolute value of the electron charge,
$ c $~is the speed of light, and
$ {\rm \bf B} $~is the magnetic field vector.

Next, let us introduce the standard definitions for the center
of mass of the system
\begin{equation}
{\rm \bf R} =  \frac{\displaystyle m_e {\rm \bf r}_e + m_i {\rm \bf r}_i }%
  {\displaystyle \vphantom{l_x^x} m_e + m_i } \: ,
\label{eq:Center_of_mass}
\end{equation}
the relative position of an electron with respect to ion
\begin{equation}
{\rm \bf r} = {\rm \bf r}_e \! - {\rm \bf r}_i \, ,
\label{eq:Relative_position}
\end{equation}
and the following mass parameters: the total mass of the system
\begin{equation}
M = m_i + m_e \approx m_i \, ,
\label{eq:Total_mass}
\end{equation}
the reduced mass
\begin{equation}
\mu = \, \frac{\displaystyle \vphantom{l_x^x} m_i \, m_e }%
      {\displaystyle \vphantom{l_x^x} m_i + m_e } \approx m_e \, ,
\end{equation}
and the relative mass difference
\begin{equation}
\nu = \, \frac{\displaystyle \vphantom{l_x^x} m_i - m_e }%
      {\displaystyle \vphantom{l_x^x} m_i + m_e } \approx 1 \, .
\end{equation}

As a result, the initial set of equations~(\ref{eq:Initial_set})
will take the form:
\begin{subequations}
\begin{eqnarray}
\frac{\displaystyle d^2}{\displaystyle d t^2} {\rm \bf R}
& = &
- \frac{\displaystyle \vphantom{l_x^x} e}%
  {\displaystyle \vphantom{l_x^x} M c} \,
  \frac{\displaystyle d {\rm \bf r}}%
  {\displaystyle d t} \! \times \! {\rm \bf B} \, ,
\label{eq:Motion_Center_of_mass}
\\
\frac{\displaystyle d^2}{\displaystyle d t^2} {\rm \bf r}
& = &
- \frac{\displaystyle \vphantom{l_x^x} e^2 }%
  {\displaystyle \vphantom{l_x^x} \mu } \,
  \frac{\displaystyle \vphantom{l_x^x} {\rm \bf r} }%
  {\displaystyle \vphantom{l_x^x} r^3 } \, -
  \frac{\displaystyle \vphantom{l_x^x} e }%
  {\displaystyle \vphantom{l_x^x} \mu c } \,
  \frac{\displaystyle d}{\displaystyle d t} ( {\rm \bf R} +
  \! \nu {\rm \bf r} ) \! \times \! {\rm \bf B} \, .
\label{eq:Motion_Relative}
\end{eqnarray}
\end{subequations}

Next, it is convenient to normalize all spatial quantities to
the initial radius~$ a $ of the unperturbed orbit:
\begin{equation}
{\rm \bf r} = a \tilde{\rm \bf r} \, ,
\quad
{\rm \bf R} = a \tilde{\rm \bf R} \, ;
\label{eq:Normalization_Spatial}
\end{equation}
and time, to the Keplerian period of revolution:
\begin{equation}
t = \tau \tilde{t} \, ,
\quad \mbox{where} \quad
\tau = 2 \pi \sqrt{{\mu a^3 \! }/{e^2}} \, .
\label{eq:Normalization_Temporal}
\end{equation}

Then, equations~(\ref{eq:Motion_Center_of_mass}),
(\ref{eq:Motion_Relative}) can be rewritten in terms of
the dimensionless variables (marked by tildes) as
\begin{subequations}
\begin{eqnarray}
\frac{\displaystyle d^2}{\displaystyle d \tilde{t}^2} \tilde{\rm \bf R}
& = &
- 2 \pi \, \frac{\displaystyle \vphantom{l_x^x} {\Omega}_M }%
{\displaystyle \vphantom{l_x^x} {\omega}_e } \,
\frac{\displaystyle \vphantom{l_x^x} d \tilde{\rm \bf r}}%
{\displaystyle d \tilde{t}} \! \times \! {\rm \bf b} \, ,
\label{eq:Motion_Center_of_mass_Dimless}
\\
\frac{\displaystyle d^2}{\displaystyle d \tilde{t}^2} \tilde{\rm \bf r}
& = &
- ( 2 \pi )^2 \frac{\displaystyle \vphantom{l_x^x} \tilde{\rm \bf r} }%
{\displaystyle \vphantom{l_x^x} \tilde{r}^3 } \, -
2 \pi \, \frac{\displaystyle \vphantom{l_x^x} {\Omega}_{\mu} }%
{\displaystyle \vphantom{l_x^x} {\omega}_e } \,
\frac{\displaystyle \vphantom{l_x^x} d}{\displaystyle \vphantom{l_x^x}
d \tilde{t}} ( \tilde{\rm \bf R} + \! \nu \tilde{\rm \bf r} )
\! \times \! {\rm \bf b} \, , \hphantom{mm}
\label{eq:Motion_Relative_Dimless}
\end{eqnarray}
\end{subequations}
where
\begin{equation}
{\Omega}_M \! = \frac{\displaystyle \vphantom{l_x^x} e B }%
                {\displaystyle \vphantom{l_x^x} M c }
\approx \,
{\Omega}_i \! = \frac{\displaystyle \vphantom{l_x^x} e B }%
                {\displaystyle \vphantom{l_x^x} m_i c }
\label{eq:Gyrofreq_Center_of_mass}
\end{equation}
is the gyrofrequency of the center of mass,
\begin{equation}
{\Omega}_{\mu} \! =
  \frac{\displaystyle \vphantom{l_x^x} e B }%
  {\displaystyle \vphantom{l_x^x} \mu \, c }
\approx \,
{\Omega}_e \! =
  \frac{\displaystyle \vphantom{l_x^x} e B }%
  {\displaystyle \vphantom{l_x^x} m_e c }
\label{eq:Gyrofreq_Relative}
\end{equation}
is the gyrofrequency of the relative motion,
\begin{equation}
{\omega}_e = {2 \pi} / {\tau} =
  \sqrt{{\displaystyle \vphantom{l_x^x} e^2 \!}/{\displaystyle \mu a^3 }}
\approx
  \sqrt{{\displaystyle \vphantom{l_x^x} e^2 \!}/{\displaystyle m_e a^3 }}
\label{eq:Keplerian_freq}
\end{equation}
is the Keplerian frequency, and
$ {\rm \bf b} \! = {\rm \bf B} / B $~is the unit vector in the direction
of the magnetic field.

Since $ m_e \ll m_i $, it can be reasonably assumed that
the center-of-mass motion is much slower than the relative motion
of the electron and ion
\footnote{
This is actually the same assumption as in the time-scale separation
method~\cite{bab06}.}:
\begin{equation}
\big| d \tilde{\rm \bf R} / d \tilde{t} \big| \ll
\big| d \tilde{\rm \bf r} / d \tilde{t} \big| \, .
\label{eq:Assumption}
\end{equation}
Consequently, the term $ d \tilde{\rm \bf R} / d \tilde{t} $
in the right-hand side of Eq.~(\ref{eq:Motion_Relative_Dimless})
can be neglected, and this equation becomes completely independent
of Eq.~(\ref{eq:Motion_Center_of_mass_Dimless}):
\begin{equation}
\frac{\displaystyle d^2 \tilde{\rm \bf r}}%
  {\displaystyle d \tilde{t}^2} \, = \,
- ( 2 \pi )^2 \frac{\displaystyle \vphantom{l_x^x} \tilde{\rm \bf r} }%
  {\displaystyle \vphantom{l_x^x} \tilde{r}^3 } \, -
  2 \pi \, \frac{\displaystyle \vphantom{l_x^x} {\Omega}_e }%
  {\displaystyle \vphantom{l_x^x} {\omega}_e } \,
  \frac{\displaystyle \vphantom{l_x^x} d \tilde{\rm \bf r} }%
  {\displaystyle \vphantom{l_x^x} d \tilde{t}}
  \! \times \! {\rm \bf b} \, .
\label{eq:Motion_Relative_Simplified}
\end{equation}
Therefore, we can solve this equation alone and then substitute
the obtained solution to Eq.~(\ref{eq:Motion_Center_of_mass_Dimless}).

Rewriting Eq.~(\ref{eq:Motion_Relative_Simplified}) in the polar
coordinate system $ (\tilde{r}, \varphi) $ leads to the following set
of equations:
\begin{equation}
\left\{ \!
\begin{array}{l}
\ddot{\tilde{r}} - \tilde{r} \dot{\varphi}^2 +
  ( 2 \pi ) \frac{\displaystyle \vphantom{l_x^x} {\Omega}_e }%
  {\displaystyle \vphantom{l_x^x} {\omega}_e } \,
  \tilde{r} \dot{\varphi} +
  ( 2 \pi )^2 \frac{\displaystyle \vphantom{l_x^x} 1 }%
  {\displaystyle \tilde{r}^2 } \, = 0 \, ,
\\[2ex]
\tilde{r} \ddot{\varphi} + 2 \dot{\tilde{r}} \dot{\varphi} -
  ( 2 \pi ) \frac{\displaystyle \vphantom{l_x^x} {\Omega}_e }%
  {\displaystyle \vphantom{l_x^x} {\omega}_e } \,
  \dot{\tilde{r}} \, = 0 \, ,
\end{array}
\right.
\label{eq:Motion_Relative_Polar_coord}
\end{equation}
where dot denotes a derivative with respect to
the dimensionless time~$ \tilde{t} $.

Let us seek for the solution of
equations~(\ref{eq:Motion_Relative_Polar_coord}) as
perturbation of the purely circular motion:
\begin{equation}
\left\{ \!
\begin{array}{l}
\tilde{r} ( \tilde{t} ) \, =
  {\tilde{r}}_0 + \delta \tilde{r} ( \tilde{t} ) \, ,
\\[1ex]
\varphi ( \tilde{t} ) =
  2 \pi \tilde{t} + {\varphi}_0 + \delta \varphi ( \tilde{t} ) \, ,
\end{array}
\right.
\label{eq:Perturbations}
\end{equation}
where $ {\tilde{r}}_0 = {\rm const}, \; {\varphi}_0 = {\rm const} $.

Then, it can be easily shown that, due to the influence of
a magnetic field, the mean orbital radius changes from~1
(in dimensionless units) to
\begin{equation}
{\tilde{r}}_0 =
  \bigg[
  1 - \frac{\displaystyle \vphantom{l_x^x} {\Omega}_e }%
  {\displaystyle \vphantom{l_x^x} {\omega}_e }
  \bigg] ^{-1/3} \, ,
\label{eq:Solution_r0}
\end{equation}
while perturbations~$ \delta \tilde{r} ( \tilde{t} ) $ and
$ \delta \varphi ( \tilde{t} ) $ experience small harmonic
oscillations with the frequency
\begin{eqnarray}
\omega & = & ( 2 \pi )
  \bigg[ \bigg(
    1 + \frac{\displaystyle \vphantom{l_x^x} 1 }%
    {\displaystyle \tilde{r}_0^3 }
  \bigg) ^{\!\! 2}
    -3 \frac{\displaystyle \vphantom{l_x^x} 1 }%
    {\displaystyle \tilde{r}_0^3 } \,
  \bigg] ^{1/2}
\! \approx
\nonumber
\\[1ex]
&& ( 2 \pi )
  \bigg[
    1 - \frac{\displaystyle \vphantom{l_x^x} 1 }%
    {\displaystyle \vphantom{l_x^x} 2 } \,
    \frac{\displaystyle \vphantom{l_x^x} {\Omega}_e }%
    {\displaystyle \vphantom{l_x^x} {\omega}_e }
  \bigg] \, , \quad \mbox{at \;}
{\Omega}_e / {\omega}_e \ll 1 \, .
\label{eq:Solution_Omega}
\end{eqnarray}

Not going into details of the perturbed relative motion,
let us mention only one property, which will be very important
further: As follows from Eq.~(\ref{eq:Solution_Omega}),
the electron \textit{after each revolution} will be shifted
by some distance \textit{approximately in the same direction}
in the course of
\begin{equation}
n \approx \frac{\displaystyle \pi }%
  {\displaystyle \vphantom{l_x^x} 2 ( 2 \pi \! - \omega ) }
  \, \approx \, \frac{\displaystyle 1 }%
  {\displaystyle \vphantom{l_x^x} 2 }
  \, \frac{\displaystyle {\omega}_e }%
  {\displaystyle \vphantom{l_x^x} {\Omega}_e }
\label{eq:Correlated_steps}
\end{equation}
revolutions; and then the direction of these shifts will gradually
change, as shown by the dotted arrow within the shaded circle
in Fig.~\ref{fig:Orbits}
\footnote{
To avoid unnecessary details, Fig.~\ref{fig:Orbits} represents,
in fact, the case $ n\,{\sim}\,1 $.}.
In other words, $ n $~gives the characteristic correlation
scale of the random walk.

Now, when the most important features of the effective
stochastic force in the right-hand side of
Eq.~(\ref{eq:Motion_Center_of_mass_Dimless}) are established,
we can proceed to studying behavior of the center of
mass~$ \tilde{\rm \bf R} $.
In general, Eq.~(\ref{eq:Motion_Center_of_mass_Dimless})
represents a stochastic differential equation, which
can be solved by various methods. In the simplest approximation,
we can use the well-known formula from the theory of random walks:
\begin{equation}
\sqrt{\big\langle {\tilde{R}}^2 (\tilde{t}) \big\rangle} \, \approx
  \Delta \tilde{R} \, \sqrt{N_{\rm tot}} \: .
\label{eq:Random_walks}
\end{equation}
Here, $ \Delta \tilde{R} $~is the typical length of one
uncorrelated step, comprising $ n $ correlated shifts
given by Eq.~(\ref{eq:Correlated_steps}); and
$ N_{\rm tot} $~is the total number of uncorrelated steps during
a given time.

Next, estimating the differential
equation~(\ref{eq:Motion_Center_of_mass_Dimless}) at the scale of
one uncorrelated step, we can easily get the following relation:
\begin{equation}
\Delta \tilde{R} \approx ( 2 \pi ) \,
  \frac{\displaystyle {\Omega}_i }{\displaystyle \vphantom{l_x^x}
  {\omega}_e } \, \Delta \tilde{r} \Delta \tilde{t} \, ,
\label{eq:One_step}
\end{equation}
where $ \Delta \tilde{r} $ and $ \Delta \tilde{t} $~are the radial
and temporal increments corresponding to such an uncorrelated step.

The last-mentioned time interval (in dimensionless units) is just
the typical number of revolutions in which the correlations still
survive, which is given by Eq.~(\ref{eq:Correlated_steps}); so that
\begin{equation}
\Delta \tilde{t} \approx n \approx \frac{\displaystyle 1 }%
  {\displaystyle \vphantom{l_x^x} 2 }
  \, \frac{\displaystyle {\omega}_e }%
  {\displaystyle \vphantom{l_x^x} {\Omega}_e } \: .
\label{eq:Correl_time_interval}
\end{equation}

On the other hand, the above-mentioned radial
increment~$ \Delta \tilde{r} $ (involving $ n $ revolutions of
the electron) is just the typical size of the shaded circle
in Fig~\ref{fig:Orbits}. Therefore, it can be estimated as
difference between the average radius of the electron orbit
in presence of the magnetic field, as given by
Eq.~(\ref{eq:Solution_r0}), and the radius without the field
(which is equal just to unity). As a result, we get
\begin{equation}
\Delta \tilde{r} \approx \, {\tilde{r}}_0 \! - 1 \, \approx
  \frac{\displaystyle 1 }{\displaystyle \vphantom{l_x^x} 3 } \,
  \frac{\displaystyle {\Omega}_e }%
  {\displaystyle \vphantom{l_x^x} {\omega}_e } \, .
\label{eq:Correl_radius}
\end{equation}

At last, the total number of random-walk steps, appearing in
Eq.~(\ref{eq:Random_walks}), can be evidently obtained as
ratio of the total time interval to the duration of one step:
\begin{equation}
N_{\rm tot} = \, \tilde{t} / n \, \approx \,
  \frac{\displaystyle {2 \Omega}_e }%
  {\displaystyle \vphantom{l_x^x} {\omega}_e } \: \tilde{t} \, .
\label{eq:Total_steps}
\end{equation}

Finally, substituting formulas~(\ref{eq:One_step})--(\ref{eq:Total_steps})
into~(\ref{eq:Random_walks}), we arrive at
\begin{equation}
\sqrt{\big\langle {\tilde{R}}^2 (\tilde{t}) \big\rangle} \, \approx \,
  \frac{\displaystyle \pi \sqrt{2} }{\displaystyle \vphantom{l_x^x} 3 } \,
  \bigg( \frac{\displaystyle {\Omega}_i }%
  {\displaystyle \vphantom{l_x^x} {\omega}_e } \! \bigg)
  \bigg( \frac{\displaystyle {\Omega}_e }%
  {\displaystyle \vphantom{l_x^x} {\omega}_e } \! \bigg) ^{\!\! 1/2}
  \tilde{t} ^{\, 1/2} \, ,
\label{eq:Diffusion_law}
\end{equation}
which is just the diffusion law, where the coefficient of diffusion
(in dimensionless units) is
\begin{equation}
\tilde{D} \approx
  \bigg( \frac{\displaystyle {\Omega}_i }%
  {\displaystyle \vphantom{l_x^x} {\omega}_e } \! \bigg) ^{\!\! 2}
  \bigg( \frac{\displaystyle {\Omega}_e }%
  {\displaystyle \vphantom{l_x^x} {\omega}_e } \! \bigg) \, .
\label{eq:Diffusion_coeff_dimless}
\end{equation}
The exact numerical factor appearing here can be hardly reliable
in the framework of our approximate treatment; so we would prefer
not to write it in the final result.

At last, returning to the dimensional variables, we get
\begin{equation}
D \approx \,
  \frac{\displaystyle e \, \sqrt{a} }{\displaystyle \sqrt{m_e} } \,
  \bigg( \frac{\displaystyle {\Omega}_i }%
  {\displaystyle \vphantom{l_x^x} {\omega}_e } \! \bigg) ^{\!\! 2}
  \bigg( \frac{\displaystyle {\Omega}_e }%
  {\displaystyle \vphantom{l_x^x} {\omega}_e } \! \bigg) \, .
\label{eq:Diffusion_coeff}
\end{equation}
This seems to be the first analytical formula derived for
the coefficient of magnetically-stimulated diffusion of
the high-angular-momentum (``circular'') Rydberg atoms.

Now, let us present some numerical estimates of the effect under
consideration. Rewriting expression~(\ref{eq:Diffusion_coeff})
in terms of the ``elementary'' physical quantities and using
relation $ a = a_0 n_q^2 $, we get
\begin{equation}
D \approx \,
  \frac{ e \, a_0^5 \, B^3 }{ m_i^2 \, c^3 } \: n_q^{10} ,
\label{eq:Diffusion_coeff_reduced}
\end{equation}
where $ a_0 $~is Bohr radius, and
$ n_q $~is the principal quantum number of the Rydberg atom.

Taking $B=$~3\,T~$= 3{\times}10^4$\,G (which can be achieved in
the installations for antihydrogen production), and
identifying~$ m_i $ with the mass of (anti-)proton, we arrive at
\begin{equation}
D \approx 7{\times}10^{-22} \, n_q^{10} \: {\rm cm}^2 \! / {\rm s} \, .
\label{eq:Diffusion_coeff_numer}
\end{equation}
Since antihydrogen atoms are typically formed by the three-body
recombination in the states with $ n_q\,{\approx}\,100$ or somewhat
greater, Eq.~(\ref{eq:Diffusion_coeff_numer}) gives
$ D{\sim}\,1 $\,cm${}^2$\,s${}^{-1}$.
Such value of the diffusion coefficient should be very
important for the experiments, because they involve trapping
of the antihydrogen atoms in a chamber of typical size about
a centimeter during the time intervals about a second.
A significant practical conclusion following from our
consideration is that it might be unreasonable to increase
the magnetic field strength above some critical value:
otherwise, the magnetically-stimulated diffusion of the created
atoms will override trapping the charged particles by the
magnetic field.

As regards ordinary ultracold plasmas, the diffusion with
coefficient $ D{\sim}\,1 $\,cm${}^2$\,s${}^{-1}$ will be
less relevant to the current experiments, because they are
usually performed at the much less time scales
(${\lesssim}\,10^{-4}$\,s). Besides, the diffusion coefficient
should be further suppressed for heavy ions by the
term~$ m_i^2 $ in the denominator of
formula~(\ref{eq:Diffusion_coeff_reduced}).
Let us emphasize also that, since the ordinary ultracold gas
density is usually not so small (${\sim}\,10^9$\,cm${}^{-3}$),
the characteristic interparticle separation will not be much
greater than the typical size of the Rydberg atoms. As a result,
the approximation of non-interacting atoms may no longer be
sufficiently adequate. (However, this approximation should
work very well for antihydrogen plasmas, whose density is
extremely low, about 1\,cm${}^{-3}$.)

At last, let us mention that, as distinct from the case of
``linear'' atoms~\cite{sch92a, sch92b}, our
formula~(\ref{eq:Diffusion_coeff}) does not show any sharp
(threshold-like) onset of the diffusion as function of the
magnetic field. This is not surprising because a distinctive
feature of the linear atoms is singularity of the electron
trajectories near the Coulomb center. Such trajectories are
specifically perturbed by the external magnetic field,
resulting in the formation of kink-like peculiarities
(\textit{cf.} Fig.~2 in Ref.~\cite{sch92a}), the intermittent
domains of various dynamics, \textit{etc}. On the other hand,
the electron trajectories of ``circular'' Rydberg atoms are
always far away from the Coulomb center and, therefore,
everything changes smoothly.

In summary, we presented a pictorial treatment and derived an
explicit analytical formula for the coefficient of
magnetically-stimulated diffusion of the high-angular-momentum
(``circular'') Rydberg atoms (in contract to the previous studies,
which were based on the numerical methods and dealt with the
``linear'' atoms). As follows from the resulting formulas, the
case of diffusion considered in our work should be especially
relevant to the Rydberg atoms formed in the experiments on
the production and trapping of antihydrogen.

\acknowledgments

A considerable part of the present work was carried out during
my visit to the Max Planck Institute for the Physics of Complex
Systems (Dresden, Germany).
I am grateful to Prof. H.~Kantz for valuable discussions and
comments.


\end{document}